\newcommand{\preprintswitch}[2]{#2} 

\documentclass{ifacconf}

\usepackage{graphicx}      
\usepackage{natbib}        
\usepackage{mathtools}
\usepackage{bm}
\usepackage{color}
\usepackage{soul}
\usepackage{siunitx}
\usepackage{booktabs}
\usepackage{amsfonts}
\usepackage{subcaption}

\graphicspath{{Bilder/}}

\begin{document}
\begin{frontmatter}

\title{System Identification for Dynamic Modeling of Large Steering Angle Vehicles\thanksref{footnoteinfo}} 

\thanks[footnoteinfo]{This work was supported as a part of NCCR Automation, a National Centre of Competence in Research, funded by the Swiss National Science Foundation (grant number 51NF40\_225155).}

\author[First]{Tobias Petri} 
\author[Second]{Simone Baratto} 
\author[Second]{Giancarlo Ferrari Trecate}

\address[First]{RWTH Aachen University, Aachen, Germany (e-mail: tobias.petri@rwth-aachen.de)}
\address[Second]{École Polytechnique Fédérale de Lausanne, DECODE lab, Lausanne, Switzerland (e-mail:\{simone.baratto, giancarlo.ferraritrecate\}@epfl.ch)}

\begin{abstract}                
This paper presents the modeling of autonomous vehicles with high maneuverability used in an experimental framework for educational purposes. Since standard bicycle models typically neglect wide steering angles, we develop modified planar bicycle models and combine them with both parametric and non-parametric identification techniques that progressively incorporate physical knowledge. The resulting models are systematically compared to evaluate the tradeoff between model accuracy and computational requirements, showing that physics-informed neural network models surpass the purely physical baseline in accuracy at lower computational cost.
\end{abstract}


\end{frontmatter}

\section{Introduction}
Autonomous driving has emerged as a transformative technology, attracting significant attention due to its potential to enhance road safety. Despite consistent progress and strong evidence supporting its safety, public acceptance of autonomous driving remains limited. For instance, the comparative study by \cite{dilillo2023comparativesafetyperformanceautonomous} demonstrated that Waymo’s autonomous service achieved markedly fewer injury and property damage claims per mile than human drivers, highlighting its superior safety performance. 
Nevertheless, societal concerns and limited exposure continue to hinder trust in this technology. Motivated by these observations, we aim at building an experimental platform where autonomous driving safety can be intuitively and convincingly demonstrated in a bumper car ride, 
showcasing collision avoidance in a playful and interactive manner. Within this setup, visitors can directly engage with the system, attempting to provoke collisions while the vehicle autonomously reacts to prevent them.

In contrast to conventional on-road scenarios, bumper car arenas present unique challenges for planning-based collision avoidance algorithms like Model Predictive Control (MPC), which require sophisticated and accurate vehicle models. Bumper cars possess exceptionally high maneuverability owing to unusually large steering angles (up to approximately 115°). This distinctive feature amplifies nonlinear dynamic effects and invalidates many typical small-angle assumptions that underpin standard vehicle models, thus motivating careful system identification and model development specific to this experimental domain.
    
A natural starting point is the family of planar bicycle models, which balance modeling fidelity and computational cost for path-planning and prediction tasks~\citep{jazar2008vehicle}. Two variants are popular in the literature: the dynamic bicycle model, which is derived from the Newton–Euler formalism and explicitly accounts for tire-force interactions, and the kinematic bicycle model, which is based on the geometric configuration of the vehicle chassis.


The potential application of the bumper car model in planning-based collision avoidance or as a digital twin in testing entails various and sometimes contradictory requirements: While accuracy and computational feasibility are central demands on the bumper car model, generalization and adaptability potentially gain importance in applications where the system is influenced by cyclically changing passengers. For this reason, we seek to explore data-driven modeling approaches that offer the potential for improved adaptability and generalization while maintaining computational efficiency and accuracy.

\textit{Related Work} \\
By considering vehicles dynamics modeling, recent works extend beyond purely first-principles modeling by incorporating data-driven identification within white/grey/black-box pipelines. 
A typical field of application of such dedicated data-driven models is autonomous racing, where the main control challenge arises
from pushing the vehicle to the handling limits. In the following paragraphs, we review relevant works that embody and advance these methodologies.

While Newton-Euler models are close to physical principles, the underlying tire model brings its highest uncertainty. Thus, the parameters of the \textit{Magic Formula} in \cite{PACEJKA20121} are estimated online using a Long Short-Term Memory (LSTM) model in~\cite{kim2022physicsembeddedneuralnetwork} and~\cite{Chrosniak_2024}. In~\cite{electronics14101935} the Newton-Euler parameters are updated online using constrained Sparse Identification of Nonlinear Dynamics (SINDy, \cite{Brunton_2016}) on recent data, allowing the autonomous system to improve trajectory tracking in an MPC-framework.
A linearization of the Newton-Euler model is combined with a Recurrent Neural Network (RNN) residual model in~\cite{CHEN2024106015}, compensating for nonlinear lateral dynamics.

Another research direction augments the kinematic bicycle model with data-driven residual models to improve predictive accuracy. Using Gaussian Process Regression (GPR) in \cite{bayesracelearningraceautonomously} and RNNs in \cite{rhode2024vehiclesingletrackmodeling} the authors achieve high-accuracy path tracking in MPC frameworks. The latter work incorporates physical information by modeling either the global motion or only the kinematic state with a neural network, showing that embedding physical laws improves accuracy while reducing network size.
\cite{ghosn2024hybridextendedbicyclesimple} proposes an extended kinematic bicycle model in which vehicle velocities are derived under inclusion of tire slip predicted by an LSTM network.

Previous works rely on assumptions that are not always met by high-maneuverable vehicles. First, their reported accuracy can degrade in regimes with wide steering range and pronounced slip, as encountered on our platform. Second, while Newton–Euler formulations can represent these regimes, their numerical stiffness and integration cost can be prohibitive in real-time applications. To the best of our knowledge, none of the existing models retains the accuracy of high steering Newton–Euler dynamics while achieving kinematic-like computational speed suitable for embedded deployment.

\textit{Contributions} \\ 
This paper addresses the above research gaps with the following contributions on system identification:

1) We tailor a Newton–Euler bicycle model to the bumper-car geometry, actuation, and large steering range, including discretization choices that enable stable multi-step prediction under low-speed. For lateral tire force we propose a partly linear function as a tradeoff between simplicity and accuracy.  

2) We adapt the extended kinematic formulation of \cite{ghosn2024hybridextendedbicyclesimple} to bumper cars with extended steering, closely tying the system state to the system actuation. On top of the kinematic backbone, we employ the SINDy framework to capture the vehicle state evolution in a physics-consistent manner and Multilayer Perceptron (MLP) neural networks to model the remaining dynamics. Furthermore, through a progressive transition from black-box to grey-box representations, we examine the advantages of embedding physical structure into the learned model.

Based of experimental data we validate the vehicle model accuracy and compare the prediction runtime of the derived models. Additionally, we compare model interpretability and its implications for physically consistent model adaption.
The results highlight that kinematic grey-box models approach or surpass the accuracy of the Newton-Euler baseline at substantially lower runtime, making them attractive for real-time use on bumper car. 


The remainder of this article is organized as follows. Section 2 introduces the modeling framework and details the adaptation of two conventional vehicle models to the dynamics of a bumper car. In Section 3, data-driven system identification is conducted based on the previously derived models. Section 4 presents the experimental data acquisition and validation. 
Finally, Section 5 summarizes the main findings and outlines potential directions for future research.

\section{Physics Based Models}
The bumper car is modeled as a rigid body on a plane surface, with its pose 
$ \bm p_k =\begin{bmatrix} p_\mathrm x& p_\mathrm y &\theta \end{bmatrix}^T_k$
defined by the two-dimensional position and the yaw angle.
The vehicle motion
$\bm v_k=\begin{bmatrix}
    v_\mathrm{x} & v_\mathrm{y} & \omega
    \end{bmatrix}^T_k$
is defined in body frame by the longitudinal and lateral velocity and the yaw rotation, all referring to the center of gravity (CoG) of the vehicle. The input provided by the driver is $ \bm u_k =\begin{bmatrix} u_s & u_m \end{bmatrix}^T_k$.

\begin{figure}[ht]
    \centering
    \def\svgwidth{0.75\linewidth}  
    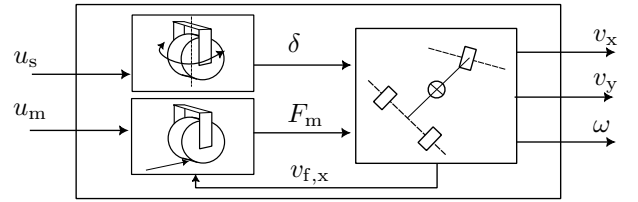    
    \caption{Block diagram of the bumper car including steering, motor, and vehicle body dynamics. }
    \label{fig:wirkungsplanCar}
\end{figure}

The bumper car consists a vehicle body with one front wheel and two rear wheels. The front wheel is both steered and motorized, while the rear axle comprises two non-driven wheels. 
Each component of the bumper car is depicted in Figure~\ref{fig:wirkungsplanCar}. 

The steering and motor dynamics are described by first-order models with saturation and a proportional motor controller\preprintswitch{\footnote{\label{fn:fullversion}Full details are reported in the extended version~\citep{petri2025systemARXIV}.}\newcounter{fnfullversion}
\setcounter{fnfullversion}{\value{footnote}}}{, detailed in Appendix \ref{app:actuators}}.

To model the dynamics of the bumper car, two widely used formulations of the bicycle model are considered in this work: (i) Newton–Euler bicycle models (often simply called dynamic bicycle model \citep{jazar2008vehicle}), which consider force-based interactions (tire slip, inertial coupling) and can achieve high fidelity, though requiring more detailed parameterization and greater computational effort; (ii) Kinematic bicycle models, which neglect tire dynamics and exploit geometric constraints for simplicity and fast computation time \citep{7225830}.

\begin{figure}[ht!]
    \centering
    \def\svgwidth{0.7\linewidth}  
\begingroup%
  \makeatletter%
  \providecommand\color[2][]{%
    \errmessage{(Inkscape) Color is used for the text in Inkscape, but the package 'color.sty' is not loaded}%
    \renewcommand\color[2][]{}%
  }%
  \providecommand\transparent[1]{%
    \errmessage{(Inkscape) Transparency is used (non-zero) for the text in Inkscape, but the package 'transparent.sty' is not loaded}%
    \renewcommand\transparent[1]{}%
  }%
  \providecommand\rotatebox[2]{#2}%
  \newcommand*\fsize{\dimexpr\f@size pt\relax}%
  \newcommand*\lineheight[1]{\fontsize{\fsize}{#1\fsize}\selectfont}%
  \ifx\svgwidth\undefined%
    \setlength{\unitlength}{384.62031bp}%
    \ifx\svgscale\undefined%
      \relax%
    \else%
      \setlength{\unitlength}{\unitlength * \real{\svgscale}}%
    \fi%
  \else%
    \setlength{\unitlength}{\svgwidth}%
  \fi%
  \global\let\svgwidth\undefined%
  \global\let\svgscale\undefined%
  \makeatother%
  \begin{picture}(1,0.76500445)%
    \lineheight{1}%
    \setlength\tabcolsep{0pt}%
    \put(0,0){\includegraphics[width=\unitlength,page=1]{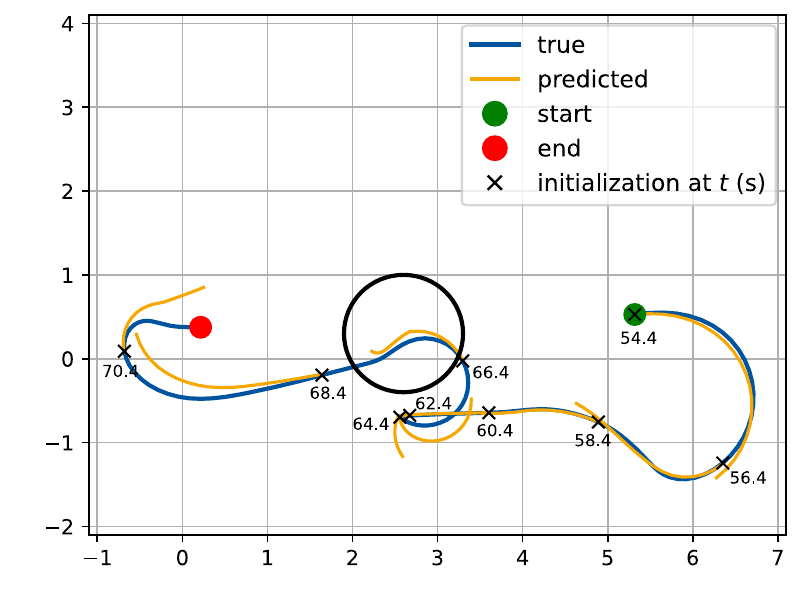}}%
    \put(0.508774,0.02659122){\color[rgb]{0,0,0}\makebox(0,0)[lt]{\lineheight{1.25}\smash{\begin{tabular}[t]{l}$p_\mathrm x$ (m)\end{tabular}}}}%
    \put(0.05711409,0.38454902){\color[rgb]{0,0,0}\rotatebox{90}{\makebox(0,0)[lt]{\lineheight{1.25}\smash{\begin{tabular}[t]{l}$p_\mathrm y$ (m)\end{tabular}}}}}%
    \put(0,0){\includegraphics[width=\unitlength,page=2]{position_plt_LNE.pdf}}%
  \end{picture}%
\endgroup%
    \caption{Comparison between the experimental vehicle trajectory and the predicted trajectory obtained with the linearized Newton-Euler model. The model is reinitialized every two seconds to test its effectiveness in different scenarios. The smaller and bigger circles isolate a specific part of the trajectories and their zoomed-out version, respectively.}
    \label{fig:original_newton_euler}
\end{figure}

Under the assumption of small steering angles, small slip angles, and approximately constant longitudinal velocities, the Newton-Euler becomes linear \citep{jazar2008vehicle}. The linear Newton-Euler model can model moderate tire slip, with moderate complexity in parametrization and evaluation. The linearization assumptions ultimately result in the diminished prediction accuracy illustrated in Figure~\ref{fig:original_newton_euler}.

The trajectory in the figure illustrates the increased maneuverability of the bumper car due to high steering angles.
Between \SI{64.4}{\s} and \SI{66.4}{\s} the steering angle exceeds~90° and the vehicle transitions between backward and forward motion, as highlighted by the orientation arrows in the zoomed-out trajectories. In these conditions the model fails to accurately predict the actual trajectory.
Further limitations result from neglecting deceleration due to tire sliding as seen between \SI{58.4}{\s} and \SI{64.4}{\s}.
These shortcomings clearly underscore the necessity for vehicle models that are reliable across a wide range of operating conditions, while preserving computational feasibility.

\begin{figure}[ht]
    \centering
    \def\svgwidth{0.9\linewidth}  
    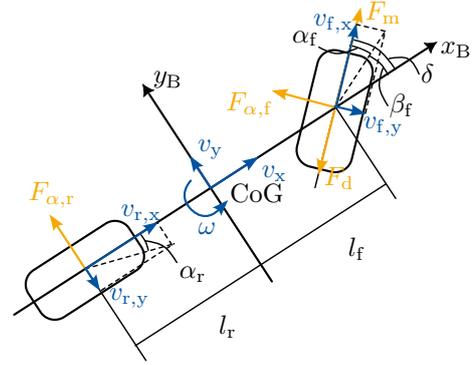    
    \caption{Newton-Euler bicycle model (velocities in blue, forces in yellow).}
    \label{fig:dynamics}
\end{figure}

\begin{figure}[ht]
    \centering
    \def\svgwidth{0.8\linewidth}  
    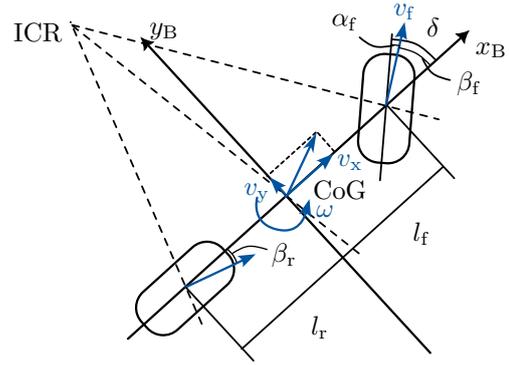    
    \caption{Kinematic bicycle model (velocities in blue). }
    \label{fig:ext_kinematics}
\end{figure}

In the following, we introduce both models and describe the modifications made to address the previously discussed limitations. The derivation and notation employed for both models follow the conventions established in~\cite{jazar2008vehicle}, specifically regarding tire slip angle $\alpha_i$, tire heading angle $\beta_i$ and wheelbase distance $l_i$ with subscript~$i \in \{f,r\}$ denoting the front (f) or rear (r) axle (cf. Figure~\ref{fig:dynamics} and~\ref{fig:ext_kinematics}). 

\textit{Newton-Euler model} \\
By applying the Newton-Euler equations of motion, one obtains 
\begin{equation}\label{eq:force balance}
    \dot{v}_\mathrm x = F_\mathrm x/{m}+\omega\, v_y,\,\,
     \dot{v}_\mathrm y = {F_\mathrm y}/{m}-\omega\, v_\mathrm x,\,\,
     \dot{\omega}={M_\mathrm z}/{I_\mathrm z} 
\end{equation}
where $m$ is the mass, $I_z$ the moment of inertia and $F_\mathrm x$, $F_\mathrm y$ and $M_\mathrm z$ the expressions for the longitudinal force, lateral force, and yaw moment.
We further derive
\begin{equation}
\begin{aligned}
   F_\mathrm x &= \sin\delta \,F_\mathrm{\alpha,f}+\cos\delta\,(F_\mathrm{m}-F_\mathrm{d})\\
   F_\mathrm y &= \sin\delta\,(F_\mathrm{m}-F_\mathrm{d})+\cos\delta \,F_\mathrm{\alpha,f}+F_\mathrm{\alpha,r}\\
   M_\mathrm z &= \sin\delta l_\mathrm f (F_\mathrm{m}-F_\mathrm{d}) + \cos\delta l_\mathrm f F_\mathrm{\alpha,f} - l_\mathrm r F_\mathrm{\alpha,r}
\end{aligned}
\end{equation}
based on the lateral slip, longitudinal drag and motor forces $F_{\alpha,i}$, $F_\mathrm{d}$ and $F_\mathrm{m}$, as illustrated in Figure~\ref{fig:dynamics}.

While longitudinal tire force is neglected at the rear wheel, we model drag-induced force at the front tire~$F_d = R_1\,v_\mathrm {f,x}+R_2\,v_\mathrm{f,x}^2$
with drag coefficients $R_1$ and $R_2$.

For a bumper car, which generally operates at low speeds and small slip angles, a linear tire force model approximates most of the dynamics, as it captures the proportional relation $F_{\mathrm{\alpha},i}\propto\alpha_i$ in that operating range. Using models with many parameters—such as the Magic Formula—is not practical because those parameters cannot be reliably identified from datasets with limited high-slip tire behavior.
To balance simplicity and realism, a suitable compromise \citep{8561285} is the following piecewise linear tire model
\begin{equation}\label{eq:latTireForce}
    F_{\alpha,i}=
    \begin{cases}
        -C_{\alpha,i} \, \alpha_i  \quad & \text{for }\quad |\alpha_i| < \hat{F}_{\alpha,i}/C_{\alpha,i}\,,\\
        -\operatorname{sign}(\alpha_i) \hat{F}_{\alpha,i} & \text{for }\quad |\alpha_i|\geq \hat{F}_{\alpha,i}/C_{\alpha,i}
    \end{cases}
\end{equation}
where $C_{\alpha,i}$ is the cornering stiffness and $\hat{F}_{\alpha,i}$ the maximum lateral tire force.
High steering angles ($|\delta|>90^\circ$) induce negative longitudinal velocity of the rear wheel. At zero-crossing the slip angle
~$\alpha_i$ loses physical meaning. To handle these situations, we adapt the Advanced Slip Ratio of \cite{lowSpeed}\preprintswitch{\footnotemark[\value{fnfullversion}]}{; additional details are reported in Appendix \ref{app:tire_slip_computation}}. Discretization ($T=\SI{0.1}{\s}$) is performed using RK4 with two subsequent steps yields the body frame velocity evolution as
\begin{equation}\label{fnewtoneuler}
    \bm v_{k+1} = \begin{bmatrix} v_\mathrm x & v_\mathrm y & \omega \end{bmatrix}_{k+1}^T = f_\mathrm{NE}(\bm v_ k,\bm u_k) \,.
\end{equation}
 
\textit{Kinematic model} \\
By using geometric relations of vehicle chassis in combination with the Instant Center of Rotation (ICR), kinematic models are widely used in path planning. In contrast to typical kinematic models, we derive a formulation with two major adaptations regarding the system representation and assumptions on the tire slip.

First, conventional kinematic models typically represent the system state using the velocity at the Center of Gravity (CoG) \citep{jazar2008vehicle, ghosn2024hybridextendedbicyclesimple}. While this choice is appropriate for vehicles exhibiting minimal tire deflection, it proves suboptimal for the bumper car, where the CoG velocity exhibits a significant correlation with the steering angle. Owing to the presence of a motorized front wheel, the front-wheel velocity $v_\mathrm{f}$ emerges as a directly actuated and thus more suitable state variable for system representation in this context.

Secondly, the assumption of slip-free movement does not hold in our context. Therefore, we derive the kinematic formulation considering the tire slip, similar to \cite{ghosn2024hybridextendedbicyclesimple}. 
The kinematic bicycle model is depicted in Figure~\ref{fig:ext_kinematics}
with the static output function $\bm v_k = g_\mathrm{kin}(\bm{x}_k)$ given by
\begin{equation}
    \begin{aligned}
        v_\mathrm x&=\cos{\beta_\mathrm f}\,v_\mathrm f\\
        v_\mathrm y&=\frac{l_\mathrm f\,\sin{\beta_\mathrm f} + l_\mathrm r\,\cos{\beta_\mathrm f}\,\tan{\beta_\mathrm r}}{l_\mathrm{f}+l_\mathrm{r}}\,v_\mathrm{f}\\
        \omega &= \frac{\sin{\beta_\mathrm{f}} - \,\cos{\beta_\mathrm{f}}\,\tan{\beta_\mathrm{r}}}{l_\mathrm{f}+l_\mathrm{r}}\,v_\mathrm{f}
    \end{aligned}\label{eq:extendedKinematicModel_f}
\end{equation}
based on the system state $\bm x_k = \begin{bmatrix}
        v_\mathrm f&\alpha_\mathrm f&\alpha_\mathrm r&\delta
    \end{bmatrix}^T$ with heading angles $\beta_\mathrm{f}=\delta+\alpha_\mathrm f$ and $\beta_\mathrm{r}=\alpha_\mathrm r$.

The evolution of~$\delta$ follows from the time discretization of the first-order steering dynamics.
The dynamics of the remaining state~$\tilde{\bm{x}}=\begin{bmatrix}
        v_\mathrm f&\alpha_\mathrm f&\alpha_\mathrm r
    \end{bmatrix}^T$ 
are not explicitly characterized from first principles. Rather, they are assumed to follow the general discrete-time model representation:
\begin{equation}\label{eq:fkin}
\tilde{\bm x}_{k+1}=\begin{bmatrix} v_\mathrm f & \alpha_\mathrm f & \alpha_\mathrm r \end{bmatrix}_{k+1}^T = f_\mathrm{kin}(\bm{x}_k, \bm{u}_k),
\end{equation}
where $f_\mathrm{kin}$ denotes a kinematic evolution function. This function is subsequently obtained via data-driven system identification, as detailed in Section~\ref{sec:datadriven}, incorporating physical insights based on motor actuation and slip dynamics.

\section{Data-Driven modeling}\label{sec:datadriven}
Despite the modifications introduced in the previous section, the models continue to exhibit inherent limitations, notably the challenge of accurately obtaining all necessary parameters through direct measurement as well as the computational requirements associated with numerical stiffness.
To overcome these limitations, data-driven methods are increasingly utilized for vehicle modeling because they can learn complex behaviors directly from data, offering faster inference and adaptability to unmodelled dynamics.

Firstly, we discuss identification of the white-box models via data-driven optimization.

Secondly, black-box approaches using MLPs are investigated. We show that these models are capable of capturing complex nonlinearities but suffer from limited physical interpretability and sensitivity to data quality, which complicates reliable online adaptation\preprintswitch{\footnotemark[\value{fnfullversion}]}{. Technical details for the implementation are reported in Appendix~\ref{app:mlp}}.

Eventually, grey-box methods are employed to merge the advantages of both approaches. By incorporating physical knowledge within a kinematic bicycle framework through SINDy, an explicit and interpretable model structure is obtained. Finally, we propose a hybrid residual modeling approach that combines the strengths of physics-based models and neural networks, enhancing prediction accuracy while preserving interpretability.


In all modeling approaches the training objective involves minimizing the Normalized Mean Squared Error (NMSE)
\begin{equation}\label{eq:lossv}
    L(\hat{\bm q},\bm q) = \frac{1}{nN}\sum_{k=1}^{N} \bigl\|\, \bm W \bigl(\hat{\bm q}_k- \bm q_k \bigr) \bigr\|_2^2
\end{equation}
with generic predicted sequence $\hat{\bm q} = \big\{ \hat{\bm q_k} \big\}_{k=1}^N, \; \hat{\bm q_k} \in \mathbb{R}^n$, measured sequence $\bm q = \big\{ \bm q_k \big\}_{k=1}^N, \; \bm q_k \in \mathbb{R}^n$ and diagonal normalization matrix $\bm W \in \mathbb{R}^{n \times n}$ to ensure consistent and comparable evaluation of prediction accuracy across models. The two sequences will be specified when discussing the individual identification methods.

\textit{White-box modeling} \\
The geometric parameters of the Newton–Euler model ($l_\mathrm f$, $l_\mathrm r$, $m$, $\hat\delta$) and parameters related to the actuators ($d$, $T_\mathrm{s}$) can be directly measured or obtained from hardware specifications. The remaining parameters are estimated with a genetic algorithm due to the highly nonlinear and partially non-differentiable dependence of the model outputs on its parameters \citep{goldberg1994genetic}. The training is performed by minimizing the NMSE on the one-step body frame velocity predictions, namely setting $\hat{\bm q} = f_\mathrm{NE}(\bm{v}, \bm{u})$ and $\bm q = \bm v $ in (\ref{eq:lossv})\preprintswitch{\footnotemark[\value{fnfullversion}]}{. The resulting parameters are reported in Appendix \ref{app:parameters}}.


\textit{Black-box modeling} \\
The first model we consider is a Nonlinear AutoRegressive model with eXogenous inputs implemented as an MLP (NARX-MLP) to capture actuators and vehicle dynamics while remaining independent of previous kinematic assumptions. In particular, the velocity evolution is described by
\begin{equation}
\bm{v}_{k+1}=f_\mathrm{NARX-MLP}\!\left(\bm{v}_ k,\ldots,\bm{v}_{k-n_y},\,\bm{u}_ k,\ldots,\bm{u}_{k-n_u}\right)
\end{equation}
with $n_\mathrm y=n_\mathrm u=2$. The training is performed by minimizing the NMSE on the one-step body frame velocity predictions, namely setting $\hat{\bm q} = f_\mathrm{NARX-MLP}(\bm{v}, \bm{u})$ and~$ \bm q = \bm v$ in (\ref{eq:lossv}).

For the second model, we approximate the kinematic state transition~\eqref{eq:fkin} with a multilayer perceptron (K-MLP): 
\begin{equation}\label{eq:f_k_mlp}
    \tilde{\bm x}_{k+1} = f_\mathrm{K-MLP}\!\left( \bm{x}_k, \bm{u}_k \right)
\end{equation}
The training is performed minimizing the NMSE on the one-step state predictions, namely setting $\hat{\bm q} = f_\mathrm{K-MLP}(\bm x,\bm u)$ and $ \bm q = \tilde{\bm x}$ in (\ref{eq:lossv}).

While these models lack physical interpretability, they provide a baseline for comparison against grey-box models, where physical knowledge of the system is explicitly incorporated to improve interpretability and potentially enhance prediction robustness.

\textit{Grey-box modeling} \\
The SINDy algorithm infers governing equations from data by representing the kinematic state transition (\ref{eq:fkin}) as a sparse linear combination of candidate nonlinear functions
\begin{equation}
    f_\mathrm{SINDy}(\bm x_k,\bm u_k)=\bm\Xi \bm\Theta^T(\bm x_k,\bm u_k)\,.
\end{equation}
The function library $\bm\Theta^T(\bm x_k,\bm u_k)$ is constructed based on available physical insight: generic polynomial terms are used when prior knowledge is limited, whereas trigonometric or physically motivated functions can be included to capture expected behaviors such as periodicity or symmetry. The coefficient vector $\bm\Xi$ contains the weights assigned to each candidate function, and sparsity constraints are applied to identify only the few terms that significantly contribute to the system dynamics \citep{desilva2020}.
The candidate libraries $\bm\Theta_i$ consisting of entries $q_{i,j}$ are derived as follows.

The front-wheel velocity is primarily affected by the motor throttle and braking, the rolling resistance, and the drag induced by tire slip. The throttle is the only energy-supplying input, active when the motor input~$u_\mathrm{m}$ exceeds the measured velocity~$v_\mathrm{f}$. Following the proportional feedback behavior of the motor controller, the candidate terms
\begin{equation}
q_{\mathrm{v},1}=\max(0,\,u_\mathrm{m}-v_\mathrm{f}), \qquad
q_{\mathrm{v},2}=\max(0,\,u_\mathrm{m}-v_\mathrm{f})\,v_\mathrm{f}
\end{equation}
are included.  
Braking effects are modeled as dissipative and proportional to negative brake input~$u_\mathrm{m}\leq0$, using
\begin{equation}
q_{\mathrm{v},3}=\min(0, u_\mathrm{m})v_\mathrm{f}, \qquad
q_{\mathrm{v},4}=\min(0, u_\mathrm{m})v_\mathrm{f}^2.
\end{equation}
Rolling resistance is represented by a third order polynomial in~$v_\mathrm{f}$ and included with the terms
\begin{equation}
q_{\mathrm{v},5}=v_\mathrm{f}, \quad
q_{\mathrm{v},6}=v_\mathrm{f}^2, \quad
q_{\mathrm{v},7}=v_\mathrm{f}^3\,.
\end{equation}
Additional resistance due to tire slip is modeled through polynomial combinations of~$v_\mathrm{f}$, $|\alpha_\mathrm{f}|$, and steering quantities $|\delta|$ and $|\dot{\delta}|$. Instances of such candidate terms are
\begin{equation}
q_{\mathrm{v},9}=v_\mathrm{f}|\alpha_\mathrm{f}|, \quad
q_{\mathrm{v},12}=v_\mathrm{f}|\dot{\delta}|, \quad
q_{\mathrm{v},16}=v_\mathrm{f}|\dot{\delta}||\alpha_\mathrm{f}|.
\end{equation}
Terms independent of $v_\mathrm{f}$ are excluded to ensure dissipative behavior, i.e., $\dot{v}_\mathrm{f}\!\to\!0$ as $v_\mathrm{f}\!\to\!0$.

The slip-angle dynamics originate from the balance between the lateral tire forces and the vehicle motion. Because slip angles vanish at standstill, each candidate term contains~$v_\mathrm{f}$ except the slip angle itself. The antisymmetric relation between turn and slip direction is enforced via absolute-value operators. This leads to the introduction of the terms
\begin{equation}
q_{\alpha,1}=\alpha_\mathrm{f},\quad
q_{\alpha,2}=v_\mathrm{f}\delta,\quad
q_{\alpha,5}=v_\mathrm{f}\dot{\delta},\quad
q_{\alpha,8}=v_\mathrm{f}\delta|\dot{\delta}|.
\end{equation}
The model obtained via SINDy regression (K-SINDy)
\begin{equation}
    \tilde{\bm{x}}_{k+1}= f_\mathrm{SINDy}\!\left( \bm{x}_k, \bm{u}_k \right)
\end{equation}
leverages the physics-informed library described above to produce interpretable representations\preprintswitch{\footnotemark[\value{fnfullversion}]}{. The resulting state equation is reported in Appendix \ref{app:sindy}}. 

With SINDy being inherently limited to the terms defined in the candidate library, its accuracy is limited in comparison to universal function approximators such as neural networks.
To bridge this gap, we propose a K-SINDy-MLP model where the K-SINDy state transition equation is augmented by a MLP 
\begin{equation}
    \tilde{\bm{x}}_{k+1}= f_\mathrm{SINDy}\!\left( \bm{x}_k, \bm{u}_k \right) +  f_\mathrm{RES-MLP}(\bm{x}_\mathrm k, \bm{u}_\mathrm k) \; .
\end{equation}
The training of the MLP is performed on the NMSE of the one-step state prediction residual error $\bm e_\mathrm{k+1}=\tilde{\bm{x}}_{k+1}-f_\mathrm{SINDy}({\bm{x}}_k,\bm{u}_k)$, namely setting $\hat{\bm q} = f_\mathrm{RES-MLP}(\bm{x}_ k, \bm{u}_ k)$ and $ \bm q = \bm e$ in (\ref{eq:lossv}).

Since the residual network only learns corrections to a meaningful analytic model, it requires fewer neurons and training iterations compared to the K-MLP model (\ref{eq:f_k_mlp}). This approach thus combines interpretability and faster execution with the accuracy of neural network approximation.

\section{Experimental validation}
Data was recorded in a rectangular arena of $7\times12\,\si{\meter}$ using an OptiTrack motion-capture system for vehicle pose $\bm p$.
The body–fixed velocities $\bm{v}$ and kinematic state $\bm x$ are not directly measured and are therefore estimated with an Extended Kalman Filter (EKF)\preprintswitch{\footnotemark[\value{fnfullversion}]}{ described in Appendix~\ref{app:stateEst}}.
We recorded one hour of driving at~\SI{10}{\hertz} and an additional sequence of \SI{110}{\s} for validation.

The comparison metrics associated with the derived models are reported in Table~\ref{tab:runtimes}.
We evaluate each model by initializing it with the starting sample of the validation dataset and driving it with the validation input sequence. By comparing the predicted and actual velocity, we assess performance in terms of the NMSE and the total computational time required for the simulation.

\begin{table}[h]
  \centering
  \caption{Comparison of derived models.}
  \label{tab:runtimes}
  \begin{tabular}{lcc} 
    \toprule
    Vehicle model & NMSE $\times 10^3$  & Computation time in ms \\
    \midrule
    Newton-Euler & 5.973 & 303.7 \\ 
    NARX-MLP  & 5.587 & 76.2 \\ 
    K-MLP & 4.594 & 85.6 \\ 
    K-SINDy & 12.960 & 37.1 \\ 
    K-SINDy-MLP & 4.765 & 81.1 \\ 
    \bottomrule
  \end{tabular}
\end{table}

The comparative analysis of all modeling approaches highlights the distinct strengths and trade-offs associated with interpretability, computational efficiency, and predictive accuracy. The Newton–Euler model provides full physical interpretability but is limited by computational stiffness, resulting in slower simulation (3.6-8.2$\times$ slower than the K-MLP and K-SINDy) and modest accuracy. On the other hand, purely black-box models such as NARX-MLP and K-MLP trade interpretability for higher expressiveness. K-MLP achieves the highest prediction accuracy with moderate computation time. The K-SINDy approach
achieves the fastest prediction time due to its analytically sparse structure and offers a degree of interpretability, but its accuracy is lower than the other models due to its limited function library. 
The hybrid K-SINDy-MLP model effectively balances these competing objectives: by integrating physical knowledge (kinematic constraints and physically motivated SINDy) with the expressiveness of data-driven residual corrections (via MLP), it maintains medium interpretability, achieves computation times suitable for real-time application while remaining comparable in accuracy to K-MLP\preprintswitch{\footnotemark[\value{fnfullversion}]}{. Additional considerations on models comparison are reported in Appendix~\ref{app:accuracy}}.

\begin{figure}[ht]
    \centering
    \begin{subfigure}[b]{\linewidth}
        \centering
        \def\svgwidth{0.7\linewidth} 
\begingroup%
  \makeatletter%
  \providecommand\color[2][]{%
    \errmessage{(Inkscape) Color is used for the text in Inkscape, but the package 'color.sty' is not loaded}%
    \renewcommand\color[2][]{}%
  }%
  \providecommand\transparent[1]{%
    \errmessage{(Inkscape) Transparency is used (non-zero) for the text in Inkscape, but the package 'transparent.sty' is not loaded}%
    \renewcommand\transparent[1]{}%
  }%
  \providecommand\rotatebox[2]{#2}%
  \newcommand*\fsize{\dimexpr\f@size pt\relax}%
  \newcommand*\lineheight[1]{\fontsize{\fsize}{#1\fsize}\selectfont}%
  \ifx\svgwidth\undefined%
    \setlength{\unitlength}{384.62031bp}%
    \ifx\svgscale\undefined%
      \relax%
    \else%
      \setlength{\unitlength}{\unitlength * \real{\svgscale}}%
    \fi%
  \else%
    \setlength{\unitlength}{\svgwidth}%
  \fi%
  \global\let\svgwidth\undefined%
  \global\let\svgscale\undefined%
  \makeatother%
  \begin{picture}(1,0.76500445)%
    \lineheight{1}%
    \setlength\tabcolsep{0pt}%
    \put(0,0){\includegraphics[width=\unitlength,page=1]{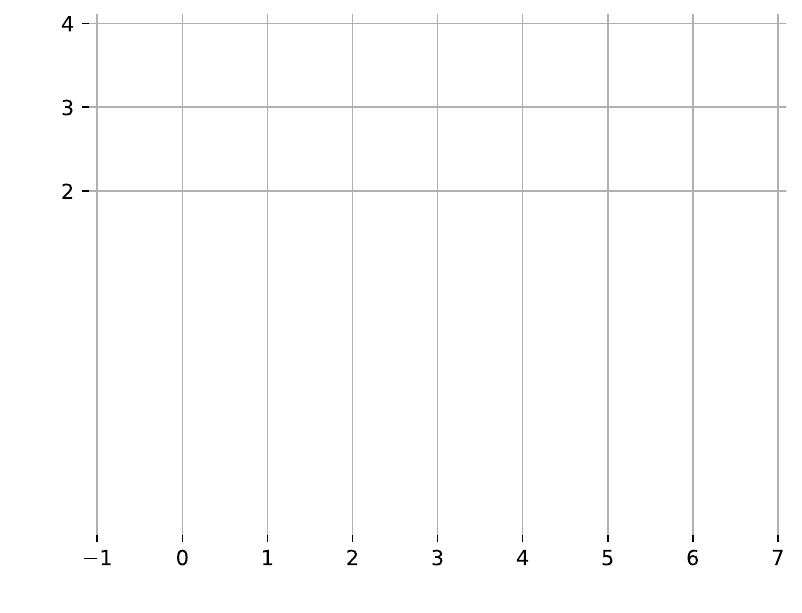}}%
    \put(0.50766119,0.02933749){\color[rgb]{0,0,0}\makebox(0,0)[lt]{\lineheight{1.25}\smash{\begin{tabular}[t]{l}$p_\mathrm x$ (m)\end{tabular}}}}%
    \put(0.05600128,0.38729526){\color[rgb]{0,0,0}\rotatebox{90}{\makebox(0,0)[lt]{\lineheight{1.25}\smash{\begin{tabular}[t]{l}$p_\mathrm y$ (m)\end{tabular}}}}}%
    \put(0,0){\includegraphics[width=\unitlength,page=2]{position_plt_NE.pdf}}%
  \end{picture}%
\endgroup%

        \caption{Newton-Euler model}
        \label{fig:position_NE}
    \end{subfigure}


    \begin{subfigure}[b]{\linewidth}
        \centering
        \def\svgwidth{0.7\linewidth} 
\begingroup%
  \makeatletter%
  \providecommand\color[2][]{%
    \errmessage{(Inkscape) Color is used for the text in Inkscape, but the package 'color.sty' is not loaded}%
    \renewcommand\color[2][]{}%
  }%
  \providecommand\transparent[1]{%
    \errmessage{(Inkscape) Transparency is used (non-zero) for the text in Inkscape, but the package 'transparent.sty' is not loaded}%
    \renewcommand\transparent[1]{}%
  }%
  \providecommand\rotatebox[2]{#2}%
  \newcommand*\fsize{\dimexpr\f@size pt\relax}%
  \newcommand*\lineheight[1]{\fontsize{\fsize}{#1\fsize}\selectfont}%
  \ifx\svgwidth\undefined%
    \setlength{\unitlength}{384.62031bp}%
    \ifx\svgscale\undefined%
      \relax%
    \else%
      \setlength{\unitlength}{\unitlength * \real{\svgscale}}%
    \fi%
  \else%
    \setlength{\unitlength}{\svgwidth}%
  \fi%
  \global\let\svgwidth\undefined%
  \global\let\svgscale\undefined%
  \makeatother%
  \begin{picture}(1,0.76500445)%
    \lineheight{1}%
    \setlength\tabcolsep{0pt}%
    \put(0,0){\includegraphics[width=\unitlength,page=1]{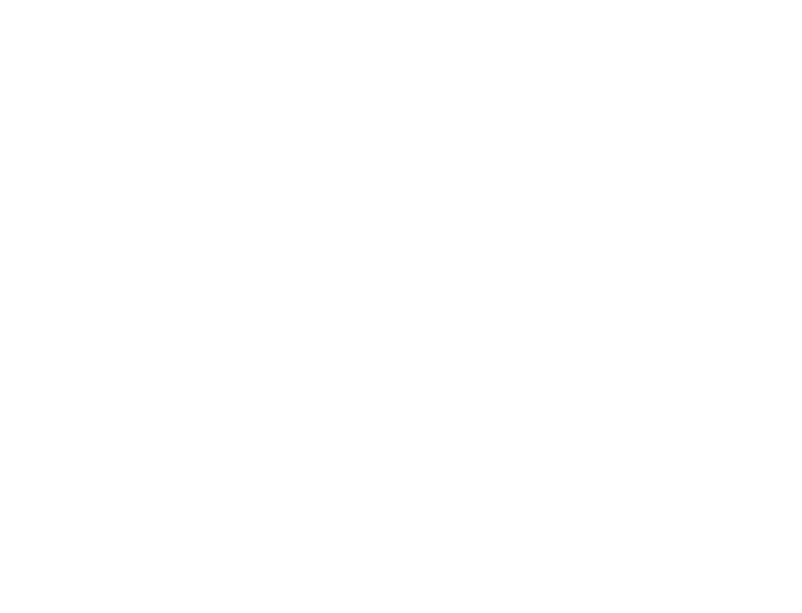}}%
    \put(0.50818987,0.02679497){\color[rgb]{0,0,0}\makebox(0,0)[lt]{\lineheight{1.25}\smash{\begin{tabular}[t]{l}$p_\mathrm x$ (m)\end{tabular}}}}%
    \put(0,0){\includegraphics[width=\unitlength,page=2]{position_plt_K-SINDy-MLP.pdf}}%
    \put(0.05652995,0.38475278){\color[rgb]{0,0,0}\rotatebox{90}{\makebox(0,0)[lt]{\lineheight{1.25}\smash{\begin{tabular}[t]{l}$p_\mathrm y$ (m)\end{tabular}}}}}%
    \put(0,0){\includegraphics[width=\unitlength,page=3]{position_plt_K-SINDy-MLP.pdf}}%
  \end{picture}%
\endgroup%
        \caption{K-SINDy-MLP model.}
        \label{fig:position_K-SINDy-MLP}
    \end{subfigure}

    \caption{Comparison between the experimental vehicle trajectory and the predicted trajectory. The models are reinitialized every two seconds to test their effectiveness in different scenarios.}
    \label{fig:overall_trajectory_comparison}
\end{figure}

To conclude this analysis, we compare the Newton–Euler and the K-SINDy-MLP models to highlight the improvements obtained through data-driven modeling. Figure~\ref{fig:position_NE} and Figure~\ref{fig:position_K-SINDy-MLP} illustrate an excerpt from the validation trajectory and the corresponding predicted trajectories. At several points along the trajectories, marked by crosses, the vehicle pose is reinitialized to mitigate integration error. A close-up view between 66.4 and 68.4 seconds captures the vehicle’s transition from backward motion ($|\delta|\!>\!90^\circ$) to forward motion ($|\delta|\!<\!90^\circ$). Within this critical segment, it is evident that the K-SINDy-MLP model substantially outperforms the Newton–Euler model (as well as its linearized version, see Figure \ref{fig:original_newton_euler}), highlighting the superior capacity of the hybrid approach to manage complex dynamical regimes.
\captionsetup[subfigure]{skip=5pt}

\section{Conclusion}
This paper investigates several modeling approaches for highly maneuverable autonomous bumper cars, comparing physical, data-driven, and hybrid frameworks. The Newton-Euler model is physically accurate but is computationally expensive. MLP models offer high accuracy and moderate runtime but lack physical interpretability, while SINDy kinematic models run fastest but with lower accuracy. Combining physical insights with neural network corrections (K-SINDy-MLP) provides a balance of speed, accuracy, and interpretability, making it well suited for real-time collision avoidance and control tasks in challenging scenarios.

Future research will concern a more refined distribution of effects in hybrid modeling. 
For instance, one could exclude deceleration terms from the base model and enforce dissipativity in the velocity residual MLP  \citep{okamoto2024learningdeepdissipativedynamics} to ensure that the correction only models additional drag rather than artificial energy injection.

\begin{ack}
The authors would like to thank Yannick Soller for his technical assistance with the experimental setup and support in collecting data on the real system.

\end{ack}

\bibliography{ifacconf}             

\preprintswitch{}
{
\newpage
\appendix

\section{Actuator dynamics}\label{app:actuators}
The stepper motor sets the steering angle $\delta$ according to the steering input $u_\mathrm{s}\in[-\hat\delta,\hat\delta]$ with steering range $\hat\delta > 0$. The steering angle dynamic is provided by a custom-made embedded controller described by the first-order model
    \begin{equation}\label{eq:deltaDot}
    \dot{\delta}=
     \begin{cases}
        ( u_\mathrm s-\delta)/T_\mathrm{s}\quad& \text{for }\quad  | u_s-\delta|/T_\mathrm s<d,  \\
        \operatorname{sign}(u_\mathrm s-\delta)d \quad& \text{for } \quad|u_\mathrm s- \delta|/T_\mathrm s\geq d
    \end{cases}
\end{equation}
with maximum steer rate~$d>0$ and time constant~$T_\mathrm{s}>0$.

The motor controller operates in throttle or braking mode depending on the sign of the motor input $u_\mathrm{m}\in[-1,c]$. 
A positive input corresponds to a reference velocity limited by $c>0$, entering a proportional controller that sets the motor force 
$F_\mathrm m = K_\mathrm t (u_\mathrm m - v_\mathrm{f,x})$, where $v_\mathrm{f,x} \in \mathbb{R}$ is the longitudinal front wheel velocity. 
Negative inputs activate braking mode, modeled as a resistive drag force~$F_\mathrm m = u_\mathrm m K_\mathrm b v_\mathrm{f,x}$.

\section{Tire Slip computation}\label{app:tire_slip_computation}
For tire slip computation, \cite{lowSpeed} introduced an Advanced Slip Ratio (ASR), where a lower bound is imposed on the longitudinal velocity when computing the tire slip angle, thus implying $\alpha\to0$ for low velocities.
Moreover, the extended steering range of a bumper cars induces negative longitudinal velocity, causing the rear wheel to rotate in reverse. As a result, the rear tire slip is given by $\alpha_\mathrm r =\beta_\mathrm r-\delta_\mathrm{r}$ where $\delta_\mathrm{r}$ is 0° if the wheel spins forward and 180° if backward. Due to the trigonometric properties of the $\arctan(\cdot)$ function, we compute the tire heading angle using the ASR and the absolute value of the longitudinal tire velocity as 
\begin{equation}\label{eq:heading angle}
    \alpha_i = \arctan\left(\frac{v_{\mathrm i,y}}{\max(|v_{\mathrm i,x}|,v_0)}\right)
\end{equation}
with longitudinal and lateral tire velocities (cf. Figure~\ref{fig:dynamics}) defined by
\begin{equation}
\begin{aligned}
    v_\mathrm{f,x}&=\cos\delta\,v_\mathrm x+\sin\delta\,(v_\mathrm y + l_\mathrm f \omega) \quad 
    && v_\mathrm{r,x}=v_\mathrm x \\
    v_\mathrm{f,y}&=\cos\delta\,(v_\mathrm y + l_\mathrm f \omega)+\sin\delta\,v_\mathrm x \quad
    && v_\mathrm{r,y}=v_\mathrm y - l_\mathrm r \omega
\end{aligned}
\end{equation}
\cite{lowSpeed} shows that the integration of ASR limits the lateral tire dynamics, however, the lateral tire dynamics remain faster than the dynamics of the vehicle body inertia, making the Newton-Euler model stiff.

\section{Hyperparameters of MLP-training}\label{app:mlp}
All models are implemented in \texttt{PyTorch} and trained to minimize the mean squared one-step prediction error. A feedforward MLP with two hidden layers of 256 and 128 neurons and ReLU activation functions is used for NARX-MLP and K-MLP, and halved number of neurons for K-SINDy-MLP. Training uses the Adam optimizer with a batch size of 256, weight decay $\lambda=10^{-5}$, and an initial learning rate of $5\cdot10^{-5}$, which is reduced by a factor of 0.6 every 40 epochs. Dropout regularization is active during training but disabled in validation. All inputs to the neural network are normalized. Each epoch consists of one gradient descent step per batch, ensuring balanced sample contribution. 

\section{Genetic algorithm for  parameter optimization of the Newton-Euler model}\label{app:parameters}
A genetic algorithm is employed for parameter identification of the Newton-Euler model due to the highly nonlinear and partially non-differentiable dependence of the model outputs on its parameters. 
The algorithm is implemented using MATLAB \texttt{ga()} function of the Global Optimization Toolbox with a population of 150 individuals and default hyperparameters.
Reasonable parameter ranges are assessed prior to optimization (for cornering stiffness $C_\alpha$ cf. \cite{tireStiffness2,tireStiffnessKlein}) and accounted for in population initialization. 

The parameters of the Newton-Euler model after optimization are given in Table~\ref{tab:parameters}.
\begin{table}[h!]
    \centering
    \caption{Parameters of the derived Newton-Euler vehicle model. Parameters are either derived from measurements, or GA-optimization (the latter is indexed by *).}
    \begin{tabular}{lccc}
        \toprule
        \textbf{Name} & \textbf{Variable} & \textbf{Value} &\textbf{Unit} \\

        \midrule
        \textbf{Steering}&&&\\
        Maximum steering angle  & $\hat\delta$ & 2.012 & rad \\
        Maximum stepper speed & $d$ & 1.418 & rad/s\\
        Stepper motor time constant  & $T_\mathrm s$ & 0.155 & s \\
                
        \midrule
        \textbf{Motor} & &&\\
        Maximum motor force* & $\hat F_\mathrm{m}$ & 423.4 &N\\
        Throttle proportional gain*  & $K_\mathrm t$ & 274.8 & Ns/m \\
        Throttle reference velocity*  & $c$ & 2.033 & m/s \\
        Brake proportional gain*  & $K_\mathrm b$ & 597.0 & Ns/m\\
       
        \midrule
        \textbf{Car Body}&&&\\
        Mass of vehicle and driver & $m$ & 319.6 & kg \\
        Moment of inertia* & $I_\mathrm z$ & 90.85 & kg$\cdot$m$^2$ \\
        Wheelbase front  & $l_\mathrm f$ & 0.54 & m \\
        Wheelbase rear & $l_\mathrm r$ & 0.33 & m \\

        \midrule
        \textbf{Tire}&&&\\
        Cornering stiffness front* & $C_\mathrm{\alpha,f} $ & 3594.0 & N/rad\\
        Cornering stiffness rear* & $C_\mathrm{\alpha, r}$ & 7840.0 & N/rad\\
        maximum front tire force* & $\hat F_\mathrm{\alpha,f}$ & 629.6 & N/rad\\
        maximum rear tire force* & $\hat F_\mathrm{\alpha,r}$ & 1360.0 & N/rad\\
        Linear front wheel drag* & $R_1$ & 43.40 & Ns/m \\
        Quadratic front wheel drag* & $R_2$ & -15.61 & N(s/m)$^2$ \\
        \bottomrule
    \end{tabular}
    \label{tab:parameters}
\end{table}

\section{SINDy regression and model parameters}\label{app:sindy}
We use the SINDy framework implemented in \textit{PySINDy} \citep{desilva2020}. All states entering the candidate matrices are scaled accordingly to ensure consistent thresholding.
 
From SINDy regression we derive the following state equations: 
\begin{subequations}
    \begin{align}
            v_{\mathrm{f},k+1}=(&0.968 v_\mathrm{f} + 0.0379 v_\mathrm{f}^2 -0.0155 v_\mathrm{f}^3 -0.475 \alpha_\mathrm{f}^2 \nonumber\\
            &+ 0.053 \max(u_\mathrm{m}c-v_\mathrm{f},0)(1+0.487v_\mathrm{f}) \nonumber\\
            &+ 0.181 \min(u_\mathrm{m} v_\mathrm{f},0)(1-0.462v_\mathrm{f}) \label{eq:sindy_vf})_k\\[0.2cm]
            \alpha_{\mathrm{f},k+1}=(& 0.777\alpha_\mathrm{f} +  0.00718 v_\mathrm{f}\delta|\delta|- 0.0117v_\mathrm{f}^2\delta\nonumber\\
            &+0.0259v_\mathrm{f}\dot\delta-0.0169v_\mathrm{f}\dot\delta|\dot\delta|-0.0169v_\mathrm{f}^2\dot\delta\nonumber\\
            &-0.00836v_\mathrm{f}\dot\delta\delta^2)_k\label{eq:sindy_af}\\[0.2cm]
            \alpha_{\mathrm{r},k+1}=(& -0.0485\alpha_\mathrm{f}  + 0.0147 v_\mathrm{f}\delta -0.0127 v_\mathrm{f}\delta|\delta|\nonumber \\
            &-0.00661 v_\mathrm{f}\dot\delta|\delta|+0.00580 v_\mathrm{f}\dot\delta\delta^2  )_k\,.\label{eq:sindy_ar}
    \end{align}
\end{subequations}

\section{State Estimation}\label{app:stateEst}
While the OptiTrack system provides the rigid-body pose $\bm p$ of the vehicle, the body–fixed velocities $\bm{v}$ are not directly measured and are therefore estimated with an Extended Kalman filter (EKF). The continuous‐time state is
\begin{equation}
\bm{x}_\mathrm{KF}=\begin{bmatrix}p_\mathrm{x}&p_\mathrm{y}&\theta&v_\mathrm{f}&\beta_\mathrm{f}&\beta_\mathrm{r}&\delta\end{bmatrix}^{T}
\end{equation}
where the first three states correspond to the measurement vector. The bumper car model with process noise $\bm{w}=[w_\mathrm v\;w_\mathrm{\beta f}\;w_\mathrm{\beta r}]^T$ is
\begin{equation}\label{eq:f_kalman_rewrite}
\dot{\bm{x}}_\mathrm{KF} = f(\bm{x}_\mathrm{KF},\bm{w}) =
\begin{bmatrix}
\cos\theta\,v_\mathrm x + \cos\theta\,v_\mathrm y\\[2pt]
\sin\theta\,v_\mathrm x - \sin\theta\,v_\mathrm y\\[2pt]
\omega\\[2pt]
w_\mathrm v\\[2pt]
\eta\,(1-\gamma)(\delta-\beta_\mathrm f) + w_\mathrm{\beta f}\\[2pt]
\eta\,(1-\gamma)(-\beta_\mathrm r) + w_\mathrm{\beta r}\\[2pt]
\mathrm{sat}_{[-d,\,d]}\!\big((u_\mathrm s-\delta)/T_\mathrm s\big)
\end{bmatrix}
\end{equation}
with $\gamma=(\tanh(10\,v_\mathrm{f}-1)+1)/2$ and $\eta=20$.
The body velocities $v_\mathrm x,v_\mathrm y,\omega$ are calculated using the kinematic model~\eqref{eq:extendedKinematicModel_f}. The term $\gamma$ drives the slip angles to zero as $v_\mathrm{f}\!\to\!0$ to avoid drift when they become unobservable; $\eta$ sets the associated pole. 
State and covariance are propagated using Runge-Kutta 4 (RK4) discretizations, following~\cite{6018993}.

\section{Discussion on predictive accuracy}\label{app:accuracy}
We compare the state prediction of the K-SINDy and K-SINDy-MLP in Figure~\ref{fig:validation}. A clear improvement in state prediction can be seen compared to the sparse kinematic model. While performance of both models is quite accurate in the first 50\si{\s}, K-SINDy-MLP excels in regimes with high-dynamic turns and rapid steering. Most prominently, slip induced deceleration at $t\approx65\si{\s}$, $t\approx78\si{\s}$ and $t\approx 97\si{\s}$ is underestimated by K-SINDy.
\begin{figure}[ht]
    \centering
    \def\svgwidth{0.99\linewidth}  
    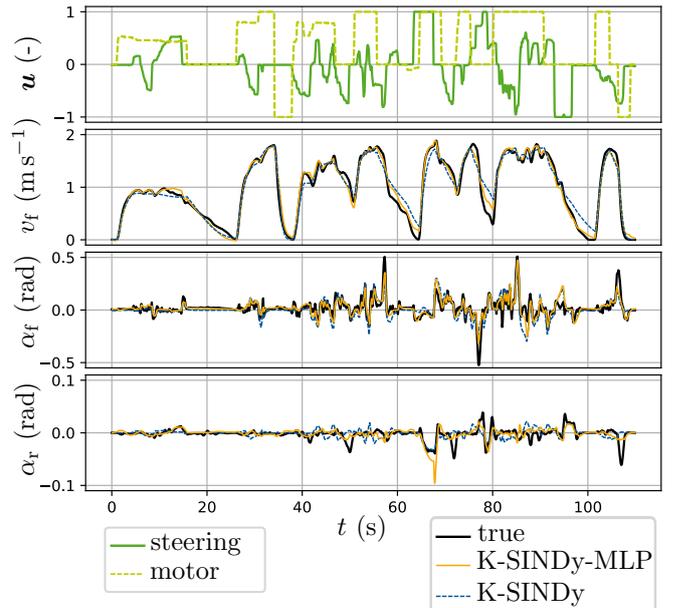    
    \caption{Closed-loop state predictions of the validation sequence of the derived K-SINDy and K-SINDy-MLP models.}
    \label{fig:validation}
\end{figure}

For the same sequence, the NARX-MLP achieves comparable fidelity but exhibits its largest deviation during backward motion at high steering at $t\approx\SI{66}{\second}$. A possible explanation can be found in the lack of kinematic constraints and internal steering dynamics. The Newton-Euler model reproduces motor impact and sliding effects closely; small under-deceleration occurs at low~$v_\mathrm f$.
}

\end{document}